# Re-Finding Found Things:
# An Exploratory Study of How Users Re-Find Information

**Robert G. Capra**
**Manuel A. Pérez-Quiñones**
Department of Computer Science
Virginia Tech
Blacksburg, VA 24061 USA
{ rcapra | perez }@vt.edu

**ABSTRACT**
The problem of how people *find* information is studied extensively; however, the problem of how people organize, re-use, and *re-find* information that they have found is not as well understood. Recently, several projects have conducted *in-situ* studies to explore how people re-find and re-use information. Here, we present results and observations from a controlled, laboratory study of re-finding information found on the web.

Our study was conducted as a collaborative exercise with pairs of participants. One participant acted as a retriever, helping the other participant re-find information by telephone. This design allowed us to gain insight into the strategies that users employed to re-find information, and into how domain artifacts and contextual information were used to aid the re-finding process. We also introduced the ability for users to add their own explicitly artifacts in the form of making annotations on the web pages they viewed.

We observe that re-finding often occurs as a two stage, iterative process in which users first attempt to locate an information source (search), and once found, begin a process to find the specific information being sought (browse). Our findings are consistent with research on waypoints; orienteering approaches to re-finding; and navigation of electronic spaces. Furthermore, we observed that annotations were utilized extensively, indicating that explicitly added context by the user can play an important role in re-finding.

## INTRODUCTION
Computer users today have many needs to store, re-find, and re-use electronic information, yet these tasks are neither well understood nor well supported by existing software tools and interfaces. People use the web to find information, but often have trouble organizing and re-finding information they have found.

A considerable amount of research is conducted to help people *find* information on the web. For example, consider research on web search engines, work in the area of recommender systems and web personalization, and research on information foraging. However, as noted by Jones, Bruce, and Dumais in a recent CIKM article about their "Keeping Found Things Found" project, the problem of "Once found, how are things organized for re-access and re-use later on?" has received "relatively less" investigation than the problem of how to find things in the first place [JBD01, p.119]. Results from a GVU study of web usage [GVU98] suggest that re-finding web pages in a problem and that users have trouble organizing information found on the web. Given that re-finding and organizing information found on the web is a problem for users, a better understanding of the re-finding process is needed to help guide the development of tools to assist users in their information re-finding tasks.

In our research, we are particularly interested in understanding and supporting the information re-finding needs and approaches of mobile users. Mobile (remote) access to information is an important dimension of re-use [JBD01] and can be especially challenging for mobile users. A recent study found that mobile workers rely heavily on the use of cell phones to enlist the help of co-workers "back at the office" to retrieve information [POS+01].

In mobile situations, users are likely to be trying to re-find information using a mobile device (PDA, cell phone) that was originally found on a different device (home or work computer). The cues and utilities of the desktop computer may not be present when the user is trying to re-access the information. We view these mobile interfaces as ways to provide directed access to satisfy specific information needs rather than as replacements for desktop applications. These directed interfaces need to support the artifacts and vocabulary that people use to communicate and reason about their re-finding needs. Thus, understanding the artifacts, vocabulary, and re-finding processes employed by users is a goal of this research project. Part of this goal stems from our interest in supporting multiple types of interfaces for mobile information access, including voice interfaces [CPR01].

Our approach is to examine re-finding from the perspective of on-going dialogs between users and their computing devices. Information found on one device may need to be referenced (re-accessed) at a later time using a different device. By investigating re-finding from a dialog perspective, we are able to gain insights into the approaches and artifacts that people use to achieve re-finding tasks. Dialog provides a good method to examine specific artifacts and their role in the re-finding process.

In particular, we show that shared context and conversational grounding are important aspects of information re-finding. Shared context are grounding are related concepts. Shared context, or common ground, refers to information that is shared among participants in a conversation [CS87]. It is information of which participants have established common understanding. Grounding refers to specific instances or points in a dialog that common understanding is reached. Examples of these concepts will be presented later in this paper.

In this paper, we present results and observations from a controlled, laboratory study we conducted to gain a better understanding of how people approach information re-finding: what they remember (recall and recognize) when trying to re-find, and how context and contextual information is used in the re-finding process.

Three key features distinguish our study from previous research on re-finding:

*Controlled, Laboratory Study* – The study was conducted as a controlled, laboratory study rather than a contextual inquiry. This allowed us to examine re-finding behaviors across identical tasks and similar situations.

*Collaborative Dialog* – We structured the study as a collaborative dialog over a telephone between two participants, one who had access to the information and one who was trying to re-find the information. This protocol was chosen because it provides good insight into the re-finding process (possibly better than talk-aloud with one participant), and also so that we could examine characteristics of the dialogs for evidence of how contextual information and shared context was used in the re-finding process. It also was chosen to support our interest in voice interfaces for information retrieval.

*Explicit Context* – We introduced the ability for participants to make and save annotations on web pages as they found information. This was done to explore if and how explicitly generated contextual information would be used in the re-finding process.

We have investigated re-finding from the perspective of how contextual information can be used to help re-locate information. Our results validate some of those found in other studies, and also contribute new insights into the use of artifacts and the approaches taken by users to re-find.

We have conducted this study as part of an effort to investigate and build better user interfaces to support mobile users needs for portable access to personal information across devices and computers.

**RELATED WORK**

Several projects have begun to explore aspects of information re-finding. We summarize these here and note a number of previous projects that have developed tools for information organization and re-use.

**Information Re-Finding**

Two current projects are closely related to our work and need special mention. Jones et al. [JBD01] and Alvarado, Teevan, Ackerman, and Karger [ATA+03] conducted *in-situ* studies of people engaging in re-finding tasks. By looking at re-finding in context, these studies have been able to examine what types of re-finding needs people have and how they use existing tools (such as email and bookmarks) to support these re-finding needs.

*Keeping Found Things Found* – Jones et al. have investigated users' behaviors and techniques for organizing and re-accessing information found on the web in their "Keeping Found Things Found" project [JBD01]. They found that people use a variety of methods for re-use and that the choice of method may depend on what function(s) are trying to be supported. For example, sending an email message to oneself with a URL is a good method for re-use if the goal is to support remote access and to serve as a reminding function. Based on their observations, Jones, et al. identified methods and functions that are important for information re-use. However, their work did not observe the dynamic process that users follow when trying to re-find information.

*Haystack* – Alvarado et al. [ATA+03] have investigated several aspects of information re-finding in a study as part of the MIT Haystack project. They conducted an inquiry in which they interrupted 15 participants in their normal work environments twice a day for five days to conduct short, directed interviews regarding their most recent information seeking activities. Among their findings, they identified two main approaches that participants took when looking for information: *orienteering* and *teleporting*. Orienteering, as defined in their paper, "involves using contextual information to narrow in on the actual information target, often in a series of steps" and is a type of situated navigation [ATA+03, p.3]. The other approach they observed was teleporting, or an attempt by the user "to take themselves directly to the information they're looking for" [ATA+03, p3]. Teleporting could be viewed as a type of plan-based navigation [JF97].

We have found a number of similar findings between their research and the results we present here, despite very different experimental approaches. We have observed both orienteering and teleporting approaches in our data. Alvarado et al. also made an important observation: "people maintained a large amount of contextual information about the specific piece of information they are looking for" [ATA+03, p.4]. We also observed that contextual information plays an important role in an iterative re-finding process.

*Waypoints* – The process by which users initially find information may have a significant impact on how they attempt to re-find it [MB97]. In a study of users performing and recalling web searches, Maglio and Barrett [MB97] observed that searchers tended to have routines for searching and that they recalled only a few important sites, or *waypoints*, on the path to their goal. Participants in their study performed web searches one day and then, on the following day, were asked to "verbally recall" [MB97, p.6] and re-create the searches. Waypoints also appear to play a reminding function. Users may be able to recall certain waypoints, but also may rely on being able to recognize information contained at waypoints to help them get further toward their goal.

*Recall and Recognition* –In a study by Mayes, Draper, McGregor, and Oatley [MDM+88], users of word processing software were not able to recall menu items from the word processor when asked to describe them on a questionnaire. Mayes et al. provided several possible explanations for why users had trouble with recall of the menu items on the questionnaire, but no trouble using them in the word processor. One of their explanations that has relevance here is that if users can rely on information to be found in the environment, they may not commit it to memory because they can rely on re-finding it in the environment when it is needed [MDM+88, p.285]. We believe this is an important concept in how people organize and plan for information re-use and how people approach the process of trying to re-find information.

*Addressability of Information* – Recall and recognition are related to the notion of what we refer to as "the addressability of information" [Ram02, p.12]. Addressability concerns how different paths, connections, access mechanisms and approaches can be used to describe the location of information. For example, for some web sites, a user may recall the specific URL for that site. In this way, the user is addressing the web site directly by its URL. However, some web sites have difficult to remember URLs, or may be accessed infrequently. In these cases, it may be easier to rely on a different form of addressing. For example, a user may rely on knowing that the CIKM website can be re-located by going to a web search engine, entering "CIKM", and browsing the top results. In this way, the site is accessed in a way that is already familiar to the user (i.e. the search engine is familiar and the search string is familiar) [Ram02].

**Tools for Re-Finding**

Information found on the web often takes the form of *semi-structured* data [Abi97] [NAM97]. Several projects at Apple Computer and one at Intel explored the use and manipulation of semi-structured pieces of information, or "information nuggets" [LNW99], contained in larger sources. These projects included Apple Data Detectors [NMW98], LiveDoc [MB98], DropZones [BM98], Grammex [LNW99], and the Intel Selection Recognition Agent [PK97].

Remembrance Agent [RS96], Margin Notes [Rho00], and Haystack [AKS99] attempt to help users collect and use personal stores of information by observing users' interactions with documents and applications. Furthermore, Remembrance Agent, Margin Notes, and Watson [BH00] use information gathered by observing users' interactions with applications to try to make recommendations of other relevant information based on the user's current tasks. A recent article by Steve Lawrence [Law00] provides an excellent survey of systems that make use of contextual information in searching web information.

The research in re-finding has not explored in a controlled, laboratory setting how users go about re-fining information. Many of the ideas explored in the literature concern the iterative process that users follow to re-find information or the information artifacts (e.g. waypoints) used in re-finding. In our study, we conducted a controlled laboratory experiment with the goal to explore further the process that users follow to re-find information and the types of information artifacts they use. The next section describes this study and the results obtained.

**RE-FINDING STUDY**

In this section, we describe the study we conducted to explore the approach users take to re-finding information. We had three goals for the study. First, we were interested in observing the fluid process that users follow to re-find previously seen information. To explore this, our study involved collaborative dialogues between two participants to observe the step-by-step process followed in re-finding. Second, we wanted to study the effect that information artifacts have in the re-finding process. Finally, we were interested in exploring if users could create information artifacts that could later be used in the re-finding process.

**Participants**

A total of 12 participants in six groups of two participated in the study. Participants were recruited from the Virginia Tech community and a majority were graduate students in Computer Science or Human Factors. Because this study examined dialog and language use, participants were required to speak North American English as their native language. All participants were familiar with web browsing.

**Sessions and Tasks**

The study consisted of two sessions that each lasted approximately one hour. In the first session, a participant (who we will refer to as the *User*) completed a set of tasks that involved *finding* information on the Internet using a web browser. The second session was scheduled about a week later and involved both the User from the first session, and a second participant (who we will refer to as the *Retriever*). In the second session, the Retriever helped the User complete tasks that involved re-finding information the User had found during the first session. Recordings were made of the sessions and the interactions between the participants. Additional details of each session are given below.

*First Session* – The first session involved only the User. In this session, the User was given a set of five tasks that involved finding information on the Internet using a web browser. The five tasks were, in order, 1) finding two movie showtimes at two theaters for three movies, 2) finding the phone numbers and addresses for four nearby restaurants, 3) finding information (event name, location, price, and hours) about three events or tourist locations for a trip to San Francisco, 4) finding names, price ranges, and phone numbers for restaurants in San Francisco for four different types of cuisine (Italian, Chinese, Thai, and American), 5) a user-defined task that allowed the user to decide a specific piece of information to look up on the web. These tasks were selected to provide a variety of directed and freeform information finding tasks.

The web browser was equipped with commercial software that allowed the User to make web annotations (such as highlighting, drawing, and notes) on web pages. Annotations became associated with that page so that whenever it was re-accessed, the annotations were re-displayed also. Each annotation could be given a classification. Three classification categories were made available by default: movies, restaurants, and travel. Users were also able to create their own classifications categories.

Prior to beginning the first session, Users were shown a video of instructions that explained the tasks and interactions that would take place in both the first and second sessions. This video described the role of both the User and the Retriever, and was shown to all participants. Participants were also given training on how to use the web annotation tools.

Users were instructed, 1) that they could make as many or as few annotations and classifications on web pages as they wished, 2) that all the web pages they browsed were being saved in a history log, and 3) that the retriever would have access to all their annotations and history during the second session to help them re-find information.

Users were given 45 minutes to work on the five tasks. Each task instruction page included a place for users to write down their findings as they completed the task. After 45 minutes, if the user had not completed the tasks, the experimenter notified the User and gave them the option of finishing the current task, up to a limit of one hour total.

*Second Session* – The second session was scheduled approximately one week after the first session and involved both the User and the Retriever. The Retriever was asked to arrive first and was shown the instruction video and given training on how to access the annotations and history log. The web annotation tools supported the retrieval of pages and annotations. Listings of pages with annotations could be viewed in a sidebar of the web browser and could be organized by web site and by classification label.

When the User arrived, they were seated in a different room from the Retriever. The User was presented with a new set of tasks that involved re-finding information that had been found during the first session. The User was given the same number of tasks as they completed in the first session; this was either four or five tasks for all participants. The re-finding tasks mirrored the finding tasks that had been given during the first session. The five tasks were, in order, 1) remember or re-find the name of a movie and re-find the earliest showtimes at two theaters, also re-find the rating for the movie, 2) remember or re-find the names of two restaurants and re-find their phone numbers, 3) re-find the names and locations of all the events or tourist activities related to San Francisco that were found during the first session, 4) re-find the names and addresses of one Italian and one Chinese restaurant in San Francisco, 5) try to re-find the information from the user-defined task from the first session. These tasks were selected so that they mirrored the finding tasks from the first session, but provided some variety in the information requested. In some cases, these tasks requested that users re-find a subset of information found in the counterpart task from the first session. In other cases, the task required re-finding the same path, but requested different (new) specific information. For example, the movie re-finding task (task 1) asked Users to find the movie rating although the rating was not asked for in the first session movie task.

The User did not have access to any of their information from the first session, but the Retriever did. The Retriever was seated at the computer that the User had used during the first session and had access to a complete history of the web pages that the User viewed on the first day. The Retriever also had access to any web annotations and classifications made by the User as they searched.

In the instructions, Users were informed that they should direct the re-finding process and not to simply "off-load" the task on the Retriever. Users placed telephone calls to the Retriever to accomplish the re-finding tasks.

**ANALYSIS**

We report here on our analysis of tasks 1-4 for all pairs of participants. Several Users did not complete the fifth task, so we have excluded it from this analysis. A total of 26 separate telephone conversations were collected for the six user-retriever pairs for tasks 1-4. Twenty-six conversations were collected instead of 24 because there were two instances in which participants made two phone calls as part of one task. In one case, the re-finding task description allowed users to break up the task into two parts if desired and one user did so. In the second case, one pair of participants was unable to complete the task on the first try, asked the experimenter a clarifying question, and decided to try the task again.

Transcriptions were made of the 26 conversations between the Users and Retrievers. These transcriptions were verified and then coded for conversational phases, instances of common ground, use of waypoints, use of annotations, specific information requests, and additional recalled and recognized items. Coding was conducted in three stages. In the first stage, one of the authors of the paper developed an initial coding scheme and completely coded the data. In a second stage, the coding scheme was explained to a second coder who then completely coded the data. Then, in a final stage, the two coders jointly coded the data a third time, reconciling their individual coding and making small adjustments to the coding scheme.

**RESULTS AND OBSERVATIONS**

In this section, we present results and observations from our study. We describe three main findings: 1) re-finding was conducted as an iterative process to find specific information targets; 2) the re-finding process relied heavily on the use of contextual information and domain artifacts to move closer to the goal; and 3) explicitly added artifacts (i.e. annotations) can be used to expand the addressability of information and generate additional shared context. Several of these findings validate results found in other studies of re-finding.

After completing our study and considering the results of Alvarado et al [ATA+03], we can summarize our current view of re-finding in a simple observation: re-finding often relies on using contextual information in an iterative process.

**Re-Finding as an Iterative Process**

We observed that users often take a two-stage approach to re-finding, first focusing on re-locating the source of the information (searching) and then, in a second stage, engaging in a process to re-locate the specific information being sought (browsing). We observed several features in the re-finding dialogues that provide evidence to support this claim: under-specified goal statements; initial navigation suggestions that focused on starting points rather than complete paths; and searching and browsing behavior often separated by points of grounding. We describe each of these in more detail below.

*Under-Specified Initial Goal Statements*

Users often supplied incomplete, or under-specified, queries at the beginning of the re-finding dialogues. In all the dialogues we observed, users provided some form of an initial goal statement in the first few utterances of the dialog. This was not a surprise because it was indicated in the instructions that users should direct the re-finding. However, these initial goal statements often left out details of the user's full re-finding need. For example, Figure 1 shows the initial turns of a dialog from a pair of participants for re-finding task 4 (San Francisco restaurants). In turn 1, the User provides an initial goal statement about finding restaurants, but omits the need to also find addresses. This aspect of the goal is not revealed to the Retriever until the source of the information has been found (turns 6 and 7). This example typifies a two-stage approach to re-finding that we saw repeated again and again by participants: first providing enough information to start the search for the source of the information, and then revealing more details about the specific information needed after the source has been found. This approach could then be iterated as needed for other sources and other information needs.

We have attempted to quantify the level of under-specification by analyzing the goal statements for completeness. In the Task 4 example shown in Figure 1, six pieces of information needed to be communicated in the goal: that the user is interested in (1) restaurants in (2) San Francisco that are (3) Chinese and (4) Italian, and that the user needs to find out their (5) names and (6) addresses. In Figure 1, the User communicated the need for five of the six pieces in their initial goal statement, resulting in a completeness rating of 83.3% (5/6). In our study, 79.2% (19 of 24) of the dialogues had initial goal statements that were under-specified in some way (less than 100% complete). A closer look at the components of the goals shows that typically the initial goal statements included a high percentage of contextual information and a lower percentage of specific information targets. For example, in Figure 1, the information that is not included in the initial goal statement is a specific target, the addresses. Across all the dialogues, the components of the initial goal statements contained approximately 75% contextual information (e.g. San Francisco, restaurant, Chinese) and 25% specific information targets (e.g. names, addresses, phone numbers).

| [1] | U: | Okay, I'm trying to find two restaurants in the San Francisco area. Uhm, one was Chinese, one was Italian. |
|---|---|---|
| [2] | R: | Okay. |
| [3] | U: | I think I filed them underneath that restaurants, uhm. |
| [4] | R: | San Francisco restaurants? |
| [5] | U: | Yeah. |
| [6] | R: | This must be Italian here, Bella Luna. |
| [7] | U: | Okay, Bella Luna. And the restaurant address? |

**Figure 1. Initial Goal and Navigation Statement**

Several points can be made about completeness of the initial goal statements. First, these tasks were performed while the User had a printed task description in front of them, so it might be expected that these measures of completeness are higher than could be expected for spontaneous re-finding tasks.  Second, although the completeness ratings are in general high, the fact that users still made incomplete goal statements for even the later tasks suggests that users approach re-finding with a stepwise methodology.  This is consistent with the observation of Alavarado et al. [ATA+03] that users may use a combination of orienteering and teleporting to re-find information being sought.

*Initial Navigation Suggestions*

In addition to initial goal statements, users also typically provided an initial navigation suggestion about how to re-locate the information being sought.  Similar to the goal statements, these navigation suggestions were often in the form of a starting location to begin the search, not complete paths describing the precise location of the information.  In some cases, the initial navigation suggestion was specific enough for the retriever to re-locate the information directly.  For example, in Figure 1, turn 3, the User provides a navigation suggestion about the annotation category that the information may be filed under, and after a short clarification, the Retriever locates the information (turn 6).  This could be considered an instance of what Alavarado et al. [ATA+03] refer to as teleporting.

However, more typically, the process of navigation was spread out over many turns and involved the use of recalling and recognizing artifacts such as waypoints, annotations, page descriptions, and navigation history.  Users appear to start the navigation process from the general information and move toward the more specific details they are seeking.

*Grounding Events Separating Stages*

The two stages of the iterative process we observed (first searching for the information source and then browsing for specific information) were often separated by instances of grounding on artifacts and contextual information.

Once again, Figure 1 provides a good, but short, example. There are several examples in Figure 1 of grounding, but here we will focus on turn 6.  At this point, the Retriever provides information to the User about the restaurant name, and indicates they have found some source of information about the restaurant.  In turn 7, the User understands that they are now at the information source.  This is the point of grounding.  Both the User and Retriever agree that they have reached the source.  At this point, the User shifts to asking for additional specific information that they believe to be located at that source, in this case, the address of the restaurant.  This example shows how the point of grounding separates the first stage of searching for the source and the second stage of requesting specific information.  We observed numerous instances where specific information requests quickly followed points of grounding on an information source.

**Reliance on Artifacts and Context**

Users and retrievers relied on a number of artifacts and contextual information as part of making progress in the iterative re-finding process.  Specifically, waypoints and annotations were used to help achieve points of grounding where the User and Retriever reached a common understanding of a web page, annotation, goal, or piece of information.  In this section, we will present results regarding the use of two such artifacts: waypoints and annotations.

*Waypoints*

We observed extensive use of *waypoints* [MB97] by users in their attempts to re-locate sources of information.  In our analysis involving waypoint usage, we have adopted a less restrictive view of waypoints than Magilo and Barrett [MB97]; we have dropped their requirement that the waypoint definitively be along the path to the goal, and instead focus on any mention of a specific node.  This allows us to consider as waypoints even nodes that Users may mis-remember as being on the path to the goal (it is a waypoint to the User).

Waypoints were used in 20 of the 26 conversations (76.9%) we observed.  The average number of waypoints per conversation was 3.46 (stdev = 4.26).  Some participant pairs made extensive use of waypoints in their re-finding while other pairs made less use of them.  In some cases, this was due to reliance on other artifacts such as annotations and descriptions of the information being sought.  However, in some dialogues, the User and Retriever managed to achieve goals without much use of either waypoints or annotations.  This was especially true for one particular pair of participants.  In many of their re-finding dialogues, no waypoints or annotations were used, but the Retriever was especially adept at locating and finding the information requested.

To investigate how waypoints were used, we classified waypoints into three main categories:  Page/Site Titles, URLs, and Page Descriptions.  Each of these is described below.

*Page and Site Titles* – This refers to full and partial names of web sites and web pages and also to names of groups or entities associated with pages and sites.  Some examples from our data include: "the Outback Steakhouse website," and "Regal Cinema site."

*URLs* – These are spoken references to URLs and were often formed as "*<name>* dot com", to refer to the top-level "home" page for a particular web site.  In many cases, these references communicate both a URL and a site title.  For example, we observed, "Fandango dot com," "W W W dot Macados dot com," and "Movie of Yahoo dot com."

*Page Descriptions* – This refers to descriptions of the contents of a web page or site.  Some examples from our data include: "it's kind of like a Yellow Pages kind of thing," and "a sort of general page listing {pause} of many different restaurants in Blacksburg and their addresses and phone numbers and such."

Both Users and Retrievers provided page descriptions, and sometimes the description became a collaborative process that helped solidify that they had reached common ground at a page. The example shown in Figure 2 illustrates a case in which the Retriever completed a page description for the User (turn 2). Based on the common ground of the page, the user then quickly made a request for specific information believed to be on the page.

Figure 3 shows the usage of the three categories of waypoints (URLs, Titles, Descriptions) for Users and Retrievers across all dialogues.

As can be seen in Figure 3, both Users and Retrievers made use of waypoints to help the re-finding process. However, Users made more references to specific URLs. URLs can be used in attempts to teleport to information, so it is not surprising to see more use of them by Users than Retrievers. Another note to be made here concerns recall versus recognition. In many cases where Users made use of a waypoint, it was something they recalled and used to help navigate to a source of information. Retrievers often presented waypoints to Users for recognition in order to help the navigation process.

| [1] | U: | Yeah, it should, like, have, like a lots of stuff up at the top, but the movies are actually, like, down… |
| [2] | R: | In front of the page. |
| [3] | U: | Yeah. |
| [4] | R: | I see a little {unintelligible} of Men In Black Two. |
| [5] | U: | And, {unintelligible} again, the earliest times. {pause}What's the name of that theater anyways? |

**Figure 2. Example of Collaborative Grounding on a Page Description**

|  | Users | | Retrievers | |
|---|---|---|---|---|
|  | # | %User | # | %Ret |
| URLs | 8 | 17.8% | 1 | 1.8% |
| Titles | 17 | 37.8% | 27 | 48.2% |
| Desc. | 20 | 44.4% | 28 | 50.0% |
| Total | 45 | 100.0% | 56 | 100.0% |

**Figure 3. Waypoint Usage by Type**

Next, we examine the use of another type of artifact used to help re-find information, annotations.

*Annotations*

As with waypoints, annotations were a type of artifact used to help re-locate sources of information. However, annotations differ from waypoints in that we added the ability for users to *explicitly* create the annotations, while waypoints were an *implicit* piece of context that are a naturally occurring aspect of web browsing. We introduced the ability for users to create annotations in order to examine if and how they would be used to help re-finding.

So that Users would understand the possible value of annotations, Users were told in the first session that the annotations they made would be available to the Retrievers during the second session. However, Users were told that they were free to make as many or as few annotations as they wished.

Annotations were referenced in 22 of the 26 conversations (84.6%) we collected. The average number of annotations per conversation was 6.83 (stdev = 7.38). To investigate how annotations were used, we classified annotation references into three main categories: Category Names (Cat), Annotation Type (Type), and general references to annotations (Ref). Each of these will be described below.

*Category Names*. Each annotation could be associated with a named category. Two examples of references to category names from our data are: 1) "User: So there should be actually a section, in my annotations, called *restaurants*", and 2) "User: And I believe if you go into my notations… If you click on the *matinee* one… it'll pull up some stuff…"

*Annotation Types*. There were several types of annotations that could be made on the web pages: text could be highlighted, text notes could be added, and drawings could be made on the pages (such as circling items or putting arrows or an "X" next to them). Users and Retrievers made references to these specific types of annotations.

*References to Annotations*. Sometimes a reference would be made not to a specific annotation type or category name, but just to the annotation feature in general. For example, "Retriever: Do you remember how you annotated it?"

Figure 4 shows the usage of the three classes of annotations (Categories, Types, and References) for Users and Retrievers across all dialogues.

|  | Users | | Retrievers | |
|---|---|---|---|---|
|  | # | %User | # | %Ret |
| Cat | 28 | 35.0% | 28 | 31.5% |
| Type | 20 | 25.0% | 45 | 50.6% |
| Ref | 32 | 40.0% | 16 | 18.0% |
| Total | 80 | 100.0% | 89 | 100.0% |

**Figure 4. Annotation Usage by Type**

Users often included references to annotations (Ref) as part of navigation suggestions (i.e. "if you go into my notations") and this may explain the higher percentage of references (Ref) by Users than Retrievers in Figure 4. Similarly, Retrievers often provided descriptions that included the annotation types they were looking at on the screen (i.e. "those are the only two circled") and this may account for their higher percentage of references to annotation types (Type).

**Artifacts to Expand Addressability and Shared Context**

Allowing Users to make annotations on the web pages was a feature that was included in our study to allow us to examine if and how explicitly added contextual information would be used in the re-finding process. Our results (as summarized in Figure 4) indicate that both Users and Retrievers did make use of annotations and that the total number of annotation references across all conversations (80+89=169) was slightly higher than the number of waypoints used (45 + 56 = 101). In addition, Users made 35% of their annotation references to categories, indicating that they had good recall of the locations of their annotations.

This data shows a that an artifact explicitly added by the User can be useful in the re-finding process and can help increase the addressability of information sources and increase shared context.

**DISCUSSION**

Based on our observations, users often take a two-stage, iterative approach to information re-finding. The first stage is a searching process with a goal to identify and reach the source of the target information. Contextual information and domain artifacts such as waypoints were used to facilitate the search process. In the collaborative dialogues that we observed, the User often expected a confirmation that a source had been reached before proceeding to make requests for specific information that was part of their goal. This matches the collaborative nature of conversations [CS87] and may be an artifact of our experimental setup. However, our results are consistent and support both Maglio and Barrett's findings on the use of waypoints [MB97] and Alavarado et al.'s findings regarding the orienteering approaches to re-finding they observed [ATA+03]. Our results are also consistent with concepts from the navigation of electronic spaces [JF97]. We contribute that reaching points of common ground appear to play an important role in the re-finding process.

Another characteristic marked this first, searching stage. Users often did not request all the desired information up front. At times, they provided very partial information, allowing the Retriever to reach a location before making the next request. Future work should explore if a search tool for re-finding should support this approach of making incremental progress towards the goal. Research could also examine how search tools can address user needs for re-finding. Alvarado et al. have suggested that users may prefer the orienteering approach over a search-engine-style keyword search [ATA+03]. It is possible that information found on the web may not be remembered by the addressability of the information, that is, not by the *path* needed to reach the information, but instead by a *process* of how to reach the information. This process may be highly collaborative in nature, requiring several mid-point confirmations that the user is on the right way. But, more importantly, the confirmations/feedback may be required to recall the next step in the process. Research on waypoints [MB97], information scent [CPC+01], recall/recognition [MDM+88], orienteering/teleporting [ATA+03], and our work seem to support this incremental process of information re-finding.

Different artifacts are used to guide the waypoint identification process. The artifacts used in our study included URLs, page titles, and general descriptions of page content. We also found that new artifacts can be added by the user and used successfully to help re-finding. In our study, these new artifacts were in form of annotations. When they are added, we found that they could be used to help in the re-finding process and increase the addressability of the information being sought. These extra cues become integrated as parts of the context that the user looks for when trying to re-find information.

From a conversational point of view, we can say that the annotations are part of the shared context between the User and the information that was already seen on the web. This has a particular significance; since the annotation was explicitly added, it is one of the only artifacts that a re-finding tool can be certain is part of the shared context with the user. The user might not have seen the URL, title, or even parts of a page, but since the user explicitly added the annotation, there is more certainty that the user was at some point aware of the annotation. However, creating annotations requires user effort. Future work would need to explore the willingness of users to invest this time versus the potential benefit in re-finding.

With the current proliferation of mobile devices (cell phones, PDAs, MP3 players, etc.), the availability of information access from these devices, and users' needs to re-access information while away from their primary computer, it is important to explore information re-finding from a mobile standpoint. To this end, we need to understand the re-finding process that users follow well enough to support it from different tools and devices. In the mobile work environment that is available today, users need to re-find information, and often from very different devices. The study described here begins to shed light on the interactive process that users follow to re-find information and the artifacts used to achieve re-finding. Our study was conducted using a protocol that separated the User during their re-finding tasks from the computer with which they first found the information. Our results show the possibility of using annotations as context for increased information addressability and possibilities for making them available remotely from a different interaction medium.

**ACKNOWLEDGMENTS**

We thank Miranda Capra for her support and assistance with data analysis, and Dr. Naren Ramakrishnan for his continued support and feedback. This work was supported in part by the National Science Foundation under Grant No. IIS-0049075, and by a grant from IBM to explore the use of VoiceXML within their WebSphere product.